\documentclass[twocolumn,amsmath,amssymb,aps,showpacs,nofootinbib]{revtex4}

\usepackage{graphicx}

\begin{document}

\date{July 6, 2006}

\title{Inclusive pentaquark and strange baryons
production\\ in $pp$ and $\Sigma p$ collisions at high energy}

\author{I.M.Narodetskii}
\email{naro@itep.ru}
\author{M.A.Trusov}
\email{trusov@itep.ru}
\author{A.I.Veselov}
\email{veselov@itep.ru} \affiliation{ITEP, Moscow, Russia}


\begin{abstract}
We calculate the cross sections for the inclusive production in
the fragmentation region of $\Theta^+(1540)$ and $\Lambda(1520)$
in $pp$ collisions and $\Lambda(1520)$ in $\Sigma p$ collisions at
high energy using the $K$- and $\pi$-meson exchange diagrams,
respectively. The contributions of these diagrams survive at
asymptotically large energies and are energy independent in this
region up to logarithmic and power corrections. We find that
inclusive $\Theta^+(1540)$ production should be at the level of 1
$\mu\mathrm{b}\times\Gamma_{\Theta KN}/\,1$~MeV. The ratio of the
$\Theta^+(1540)$ over the $\Lambda(1520)$ yields is found to be
$\sim 1\%$. The fraction of $\Lambda(1520)$ yields in $\Sigma p$
and $pp$ collisions is $\sim 2.7$ that quantitatively agrees with
the preliminary result of the Fermilab fixed target experiment
E781.
\end{abstract}

\maketitle


\section{Introduction}

The possible existence of the $\Theta^+$ pentaquark remains one of
the puzzling mysteries  of recent years. To date there are more
than 20 experiments with evidence for this state, but criticism
for the $\Theta^+$ claim arises because similar number of high
energy experiments did not find any evidence for the $\Theta^+$,
even though the other  ``conventional'' three-quark hyperons  such
as $\Lambda(1520)$ hyperon resonance are seen clearly.

The situation is getting more intriguing, as recently CLAS
collaboration reported negative results on $\Theta^+$
photoproduction off proton and deuteron with high statistics.
Meanwhile LEPS collaboration reported the new evidence of the
$\Theta^+$ in $\gamma d\to\Theta^+\Lambda(1520)$~\cite{jlab}.
Also, DIANA collaboration using increased statistics confirms its
earlier result, and offers a new evidence for formation of a
pentaquark baryon in the charge-exchange reaction $K^+n\to K^0p$
on a bound neutron \cite{diana}. New experiments are needed to
confirm or refute the pentaquark existence.

Most of negative high energy experiments are high statistic hadron
beam experiments. \textit{E.g.} HERA-B, a fixed target experiment
at the 920 GeV proton storage ring of DESY \cite{HERA-B} finds no
evidence for narrow signals in the $\bar K^0_Sp$ channel and only
sets modest upper limits for $\Theta^+$ production of less than
16~$\mu$b/N and less than about 12$\%$ relative to $\Lambda(1520)$
in mid-rapidity region. This negative result would present serious
rebuttal evidence to worry about. However, without obvious
production mechanism of the $\Theta^+$ (if it exists) or even
$\Lambda(1520)$ the rebuttal is not very convincing.

In this paper we estimate the high-energy behavior of the
$\Theta^+$ and $\Lambda(1520)$ production cross sections in
inclusive $pp$ collisions  using the $K$ exchange diagram, which
is known to survive at high energies in the beam/target
fragmentation region. We show that the cross section of the
$\Theta^+$-production is suppressed compared to the production of
$\Lambda(1520)$. This suppression is mainly due to the smallness
of the coupling constant $G^2_{\Theta KN}$ compared  to
$G^2_{\Lambda KN}$ that in turn is related to the small width of
the $\Theta^+$. As a byproduct we also estimate the contribution
of the $\pi$ exchange diagram for the inclusive $\Lambda(1520)$
production in $\Sigma p$ collisions.

\section{Inclusive cross sections}

\begin{figure}
\includegraphics[width=50mm,keepaspectratio=true]{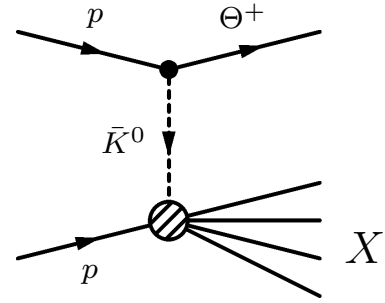}
\caption{The $\bar{K}_0$ exchange diagram for the $\Theta^+(1540)$
production in inclusive pp scattering \label{fig:pp_theta}}
\end{figure}

We assume that the $\Theta^+$ exists. Consider the $\Theta^+$
production in the reaction
\begin{equation}
\label{semiinclusive} p+p\to\Theta^++X \end{equation} where $X$ is
unspecified inclusive final state carrying the strangeness -1. The
${\bar K^0}$ exchange diagram for $pp\to\Theta^+X$ is shown in
Fig. \ref{fig:pp_theta}.

The standard expression for the single inclusive $\Theta^+$
hadroproduction cross section in $p-p$ collisions in terms of the
4-momentum transfer squared $t=q^2$ and the invariant mass
$W=\sqrt{s_1}$ of the $\bar K^0p$ system is well known (see
\textit{e.g.}  \cite{Yao}). At high energy, it is more convenient
to convert this expression to an integral over the Feynman
variable $x_F$, the fraction of the incident proton momentum
carried by the $\Theta^+$ in the initial direction of the proton
(in the center-of-mass system), and $k_{\bot}$, the transverse
momentum of $\Theta^+$ relative to the initial proton direction.
Then the contribution of the $K$ meson exchange to the double
differential cross section for the $\Theta^+$ inclusive production
reads
\begin{equation}
\label{dsigma}
\frac{d\sigma}{dx_Fdk_{\bot}^2}=\frac{1}{4\pi}\,\frac{G_{\Theta
KN}^2}{4\pi}\, J\, \frac{p}{E_{\Theta}}\,
\Phi_{p\,\Theta}(t)\,F^4(t)\, \sigma_{\text{tot}}^{\bar
K^0p}(s_1),
\end{equation}
where
\begin{equation}E_{\Theta}=\sqrt{x_F^2p^2+k_{\bot}^2+m_{\Theta}^2},
\end{equation}
\begin{equation}
s_1=s(p)+m_{\Theta}^2-2E_{\Theta}\sqrt{s(p}),\end{equation}
${s(p)}=4(p^2+m_p^2)$ being the center-of-mass energy squared, and
\begin{equation}
t=m_p^2+m_{\Theta}^2-2E_{\Theta}\sqrt{p^2+m_p^2}+2x_Fp^2 .
\end{equation}
The factor \begin{equation} J=\sqrt{\frac{\lambda(s_1,q^2,m_p^2)}
{\lambda(s,m_p^2,m_p^2)}},
\end{equation} where
\begin{equation}
\lambda(s, m_1^2,m_2^2)=s^2-2s(m_1^2+m_2^2)+(m_1^2-m_2^2)^2,
\end{equation}
is the ratio of flux factors in the $pp$ and $\bar K^0p$
reactions. The function $\Phi_{p\,\Theta}(t)$ is the squared
product of the vertex function for $p\to\Theta^+\bar K^0$ and the
kaon propagator:
\begin{equation}
\label{phi}
\Phi_{p\,\Theta}(t)=\frac{(m_p-m_{\Theta})^2-t}{(t-m_K^2)^2}.
\end{equation}
To evaluate the cross sections away from the pole position
$t=M_{K}^2$ we include  the phenomenological form factor
$F_{K}(t)$.

\begin{figure}
\includegraphics[width=50mm,keepaspectratio=true]{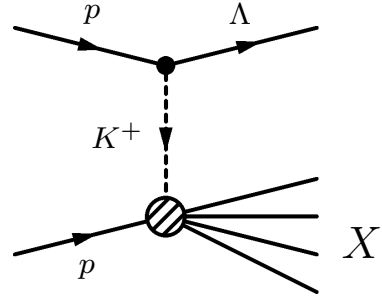}
\caption{The $K^+$ exchange diagram for the $\Lambda(1520)$
production in inclusive pp scattering \label{fig:pp_lambda}}
\end{figure}

In the high energy limit with accuracy $\mathcal{O}(1/p^2)$
\begin{equation}\label{hel}
\begin{gathered}
J\cdot \frac{p}{E_{\Theta}}\approx\frac{1-x_F}{x_F},\quad
s_1\approx (1-x_F)s,\\ t\approx
m_{\Theta}^2+m_p^2(1-x_F)-\frac{m_{\Theta}^2+k_{\bot}^2}{x_F},
\end{gathered}
\end{equation}
and Eq. (\ref{dsigma})  written in terms  $x_F$ and $k_{\bot}^2$
reads
\begin{equation}
\label{dsigma_he} \frac{d\sigma}{dx_Fdk_{\bot}^2}
=\frac{1}{4\pi}\frac{G_{\Theta KN}^2}{4\pi}\cdot \frac{1-x_F}{x_F}
\Phi_{p\,\Theta}(t)F^4(t)\sigma_{\text{tot}}^{{\bar K}^0 p}(s_1).
\end{equation}
 Eq. (\ref{dsigma_he}) can be generalized
for the fragmentation of a baryon $a$ into a baryon $b$ due to the
exchange by the meson $m$ (in particular, for the $\Lambda(1520)$
production in the reaction $p+p\to\Lambda(1520)+X$, see Fig.
\ref{fig:pp_lambda}):
\begin{equation}
\label{eq:dsigma_gen} \frac{d\sigma_{ab}}{dx_Fdk_{\bot}^2}
=\frac{1}{4\pi}\,\frac{G^2_{bma}}{4\pi}\, \frac{1-x_F}{x_F}\,
\Phi_{ab}(t)\,F^4(t)\,\sigma_{\text{tot}}^{mp}(s_1).
\end{equation}
Explicit values of the coupling constants $G^2_{bma}/4\pi$ and
expressions for $\Phi_{ab}$ used in our calculations are given
below.

\subsection{The $\Theta^+KN$ vertex}

\noindent The $\Theta^+ KN$ vertex for
$J^P(\Theta^+)={\frac{1}{2}}^+$ is
\begin{equation}
\label{NKTvertex} { L}_{\Theta KN}=iG_{\Theta
KN}(K^{\dag}{\bar\Theta}\gamma_5N+\bar N\gamma_5\Theta K),
\end{equation} with the operator $\gamma_5$ corresponding to
positive $\Theta^+$ parity. The Lagrangian (\ref{NKTvertex})
corresponds with the $\Theta^+$ being a $p$-wave resonance in the
$K^0p$ system. The partial decay width $\Gamma_{\Theta\to  K^0p}$
is
\begin{equation}\label{NKTwidth}\Gamma_{\Theta\to K^0p}=\frac{G^2_{\Theta
KN}}{4\pi}\cdot\frac{2p_K^3}{(m_{\Theta}+m_p)^2-m_K^2} ,
\end{equation} where $p_K=260$ MeV/c is the kaon momentum in the
rest frame of $\Theta^+$. To extract the value for $G_{\Theta
KN}$, we need the experimental information of the width
$\Gamma_{\Theta K N}$, which is not known precisely but whose
measurement is the subject of several planned dedicated
experiments, see \textit{e.g.} \cite{halla}. To provide numerical
estimates, we will use the value $\Gamma_{\Theta \to  K^0
p}=1$~MeV. This corresponds to the full width $\Gamma_{\Theta
KN}=\Gamma_{\Theta\to K^0 p}+\Gamma_{\Theta\to K^+ n}=2$~MeV,
which is consistent with the upper limit for the width derived
from elastic $KN$ scattering~\cite{Cahn:2003wq}. Evaluating
Eq.~(\ref{NKTwidth}) with $\Gamma_{\Theta \to  K^0p}=1$~MeV, we
extract the value $G_{\Theta KN}$
\begin{equation}
\frac{G_{\Theta KN}^2}{4\pi}=0.167\cdot\frac{ \Gamma_{\Theta\to
K^0p}}{1~\text{MeV}}, \end{equation}
 which will be used in
the subsequent estimates for the inclusive cross section.

\subsection{$\Lambda(1520)\,KN$ vertex}

\noindent The $\Lambda(1520)\,KN$ vertex is
\begin{equation}
\label{KNL}
 L_{\Lambda KN}=\frac{G_{\Lambda KN}}{m_K}\left(\bar
\Lambda^{\mu}\gamma_5N\partial_{\mu}K+\bar
N\gamma_5\Lambda^{\mu}\partial_{\mu}K^{\dag}\right),
\end{equation} where $\Lambda^{\mu}$ is the vector spinor for
the spin 3/2 particle. The Lagrangian  (\ref{KNL}) corresponds
with the $\Lambda(1520)$ being a $D$-wave resonance in the $K^-p$
system.

We treat $\Lambda(1520)$ relativistically using the
Rarita-Schwinger vector-spinor formalism \cite{piling} with the
density matrix
\begin{multline}
L^{\mu\nu}=\frac{1}{4}({\hat p}_{\Lambda }+m_{\Lambda
})\\{}\times\left[-g^{\mu\nu}+\frac{1}{3}
\gamma^{\mu}\gamma^{\nu}+\frac{1}{3m_{\Lambda
}}(\gamma^{\mu}p^{\nu}_{\Lambda }-\gamma^{\nu}p^ {\mu}_ {\Lambda
})+\frac{2}{3m^2_{\Lambda }}p_{\Lambda }^{\mu}p_{\Lambda
}^{\nu}\right].
\end{multline}
The  $\Lambda(1520)\to pK^-$ width is
\begin{equation}\Gamma_{\Lambda \to K^-p}=\frac{G_{\Lambda KN}^2}{4\pi}\cdot
\frac{2p_K^5}{3m_K^2}\cdot\frac{1}{(m_{\Lambda}+m_p)^2-m_K^2},
 \end{equation} where $p_K=246~\mathrm{MeV/c}$ is the
kaon momentum in the rest frame of $\Lambda(1520)$. Using the PDG
values of $\Gamma_{\text{tot}}(\Lambda(1520))=15.6$ MeV and
$\text{Br}(\Lambda(1520)\to N\bar K)=45\%$ we obtain
\begin{multline}
\Gamma_{\Lambda \to K^-p}=\frac{1}{2}\cdot \text{Br}(\Lambda \to
N\bar K)\cdot\Gamma_{\text{tot}}\\{}=\frac{1}{2}\cdot 0.45\cdot
15.6~\text{MeV}=3.51~\text{MeV}\end{multline} and
\begin{equation}
\frac{G_{\Lambda  KN}^2}{4\pi}\approx 8.14.
\end{equation}
The function $\Phi_{p\,\Lambda}(t)$ in Eq. (\ref{eq:dsigma_gen})
is
\begin{equation}\label{eq:phi-p-lambda} \Phi_{p\,\Lambda}(t)
=\frac{(m_p+m_{\Lambda })^2-t}{6m_{\Lambda }^2m_K^2}\cdot
\frac{((m_p-m_{\Lambda })^2-t)^2}{(t-m_K^2)^2}.\end{equation} The
expression (\ref{eq:phi-p-lambda}) includes the factor $1/m_K$ in
(\ref{KNL}).

\subsection{$ \Lambda(1520)\,\pi\Sigma$ vertex}

\noindent The $\Lambda(1520)\,\pi\Sigma$ vertex is written in
analogy with (\ref{KNL}):
\begin{equation}
\label{pisigmaL}
L_{\Lambda\pi\Sigma}=\frac{G_{\Lambda\pi\Sigma}}{m_{\pi}}\left(\bar
\Lambda^{\mu}\gamma_5\Sigma\partial_{\mu}\pi+\bar
\Sigma\gamma_5\Lambda^{\mu}\partial_{\mu}\pi^{\dag}\right),
\end{equation}
The  $\Lambda(1520)\to \Sigma^+\pi^-$ width is
\begin{equation}\Gamma_{\Lambda\to \pi^-\Sigma^+}=\frac{G_{\Lambda\pi\Sigma}^2}{4\pi}\cdot
\frac{2p_{\pi}^5}{3m_{\pi}^2}\cdot\frac{1}{(m_{\Lambda}+m_{\Sigma})^2-m_{\pi}^2},
 \end{equation} where $p_{\pi}=266~\mathrm{MeV/c}$ is the
pion momentum in the rest frame of $\Lambda(1520)$. Using the PDG
values of $\text{Br}(\Lambda(1520)\to \Sigma\pi)=42\%$ we obtain
\begin{multline}
\Gamma_{\Lambda\to \pi^-\Sigma^+}=\frac{1}{3}\cdot
\text{Br}(\Lambda\to
\pi\Sigma)\cdot\Gamma_{\text{tot}}\\{}=\frac{1}{3}\cdot 0.42\cdot
15.6~\text{MeV}=2.18~\text{MeV}\end{multline} and
\begin{equation}
\label{glspi} \frac{G_{\Lambda\pi\Sigma}^2}{4\pi}\approx 0.353
\end{equation}
The function $\Phi_{\Sigma\,\Lambda}(t)$ is obtained from that
given in Eq. (\ref{eq:phi-p-lambda}) by substitution $m_p\to
m_{\Sigma}$, $m_K\to m_{\pi}$.

\section{Results}

\begin{figure}
\includegraphics[width=80mm,keepaspectratio=true]{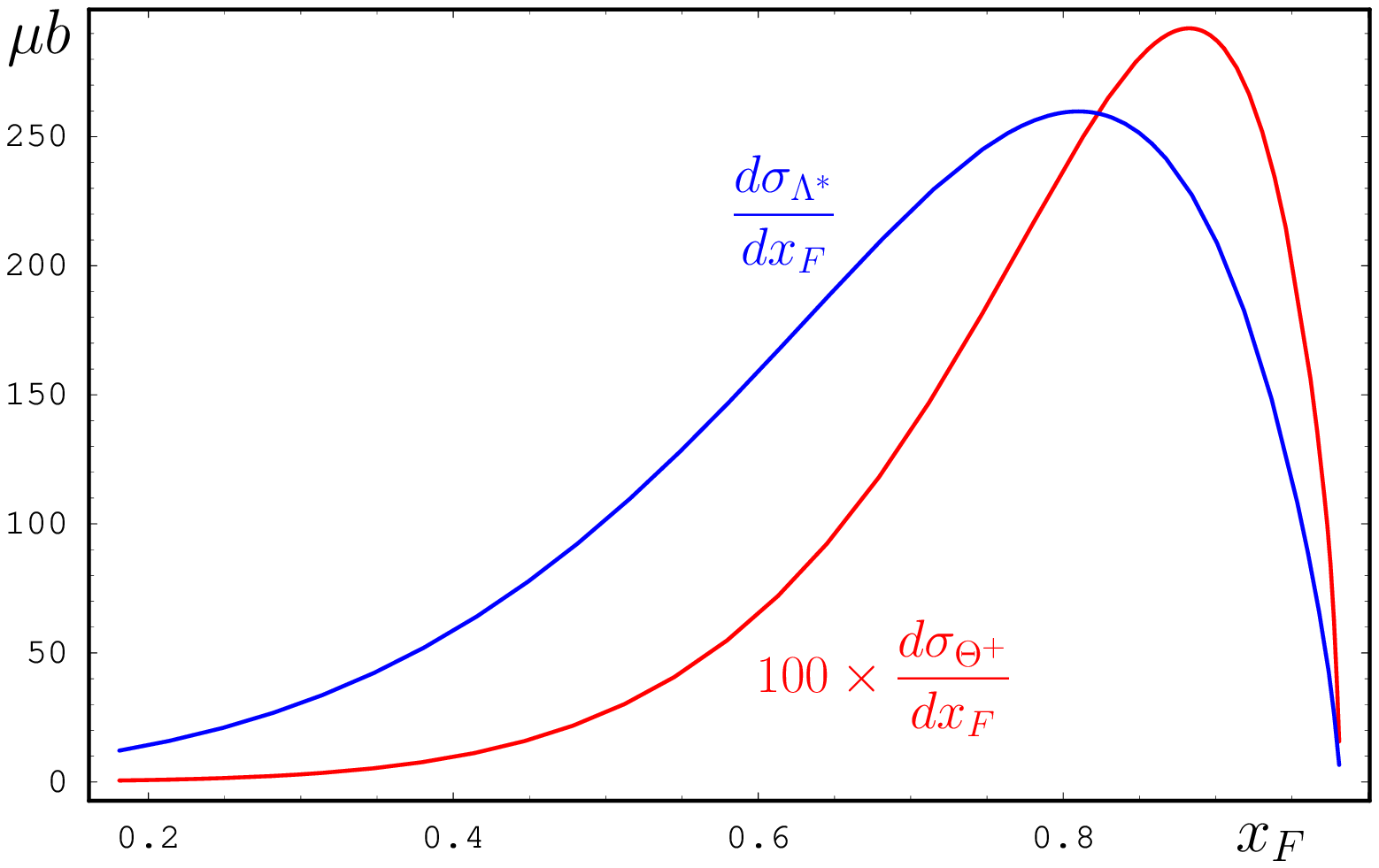}
\caption{$x_F$  dependence of the inclusive $pp\to \Theta^+(1540)$
and $\Lambda(1520)$ cross sections \label{fig:pp_inclusive:x}}
\end{figure}

\begin{figure}
\includegraphics[width=80mm,keepaspectratio=true]{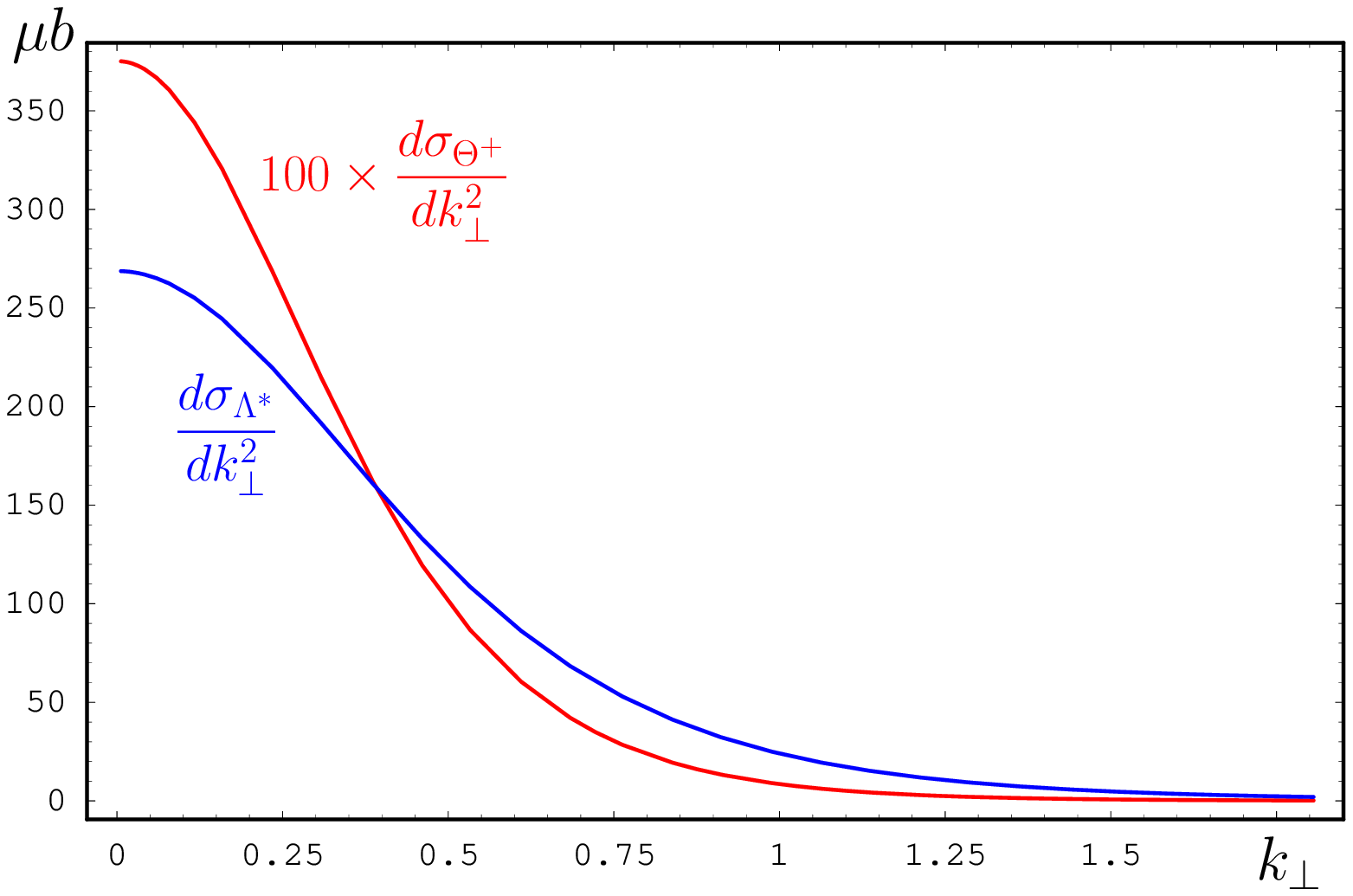}
\caption{ $k_{\perp}$ dependence of the inclusive $pp\to
\Theta^+(1540)$ and $\Lambda(1520)$ cross sections
\label{fig:pp_inclusive:k}}
\end{figure}

The total cross sections can be obtained by integrating
(\ref{dsigma_he}) over $k_{\bot}^2$ and $x_F$
\begin{equation}
\sigma_{ab}\,=\,\frac{G^2_{bma}}{4\pi}\,\int dx_F\int
dk^2_{\bot}\,K_{ab}(x_F,k^2_{\bot})\,\sigma_{\text{tot}}^{mp}(s_1),
\end{equation} where $s_1$ is given in (\ref{hel}) and
\begin{equation}
K_{ab}(x_F,k^2_{\bot})\,=\,\frac{1-x_F}{x_F}\,\Phi_{ab}(t)\,F^4(t).
\end{equation}
 The kinematical limits of integration in (\ref{dsigma_he}) are
given by equation
\begin{equation}
x_F^2+\frac{k_{\bot}^2}{p^2}\le\frac{(s(p)+m_{b}^2-s_0)^2-4s(p)m_{b}^2}{4p^2
s(p)},
\end{equation}
where $s_0=(m_p+m_K)^2$ for the $K$-exchange and
$s_0=(m_p+m_{\pi})^2$ for the $\pi$-exchange. We employ two
representative examples for the form factor $F(t)$ in
(\ref{dsigma_he}):
\begin{equation}
\mathrm{A:}~~F(t)=\frac{\Lambda^2-m_K^2}{\Lambda^2-t},~~\text{and}~~
\mathrm{B:}~~F(t)=\frac{\Lambda^4}{\Lambda^4+(t-m_K^2)^2},
\end{equation}
the cut-off parameter $\Lambda$ being a typical hadronic scale
$\Lambda=1$~GeV.

Because of (\ref{hel}) all the energy dependence of the right hand
side of Eq. (\ref{dsigma_he}) is due to the factor $\sigma_{{\bar
K}^0p}(s_1)$. Since $\sigma(s_1)$ is slow varying function of
$s_1=(1-x_F)s$, we can take it out of the integral at the point
$\hat s_1=(1-\hat x_F)s$, where $\hat x_F$ is the point at which
$d\sigma^{ab}/dx_F$ reaches the maximum\footnote{The typical
values of $\hat x_F$ are $\sim 0.8-0.9$, depending on the reaction
considered; they tabulated in Table \ref{table:results}.
Therefore, in the fragmentation region, the effective energy
$\sqrt{\hat s_1}$ is \textit{always much smaller} than
$\sqrt{s}$.}. Then we obtain
\begin{equation} \sigma_{ab} \approx \frac{G^2_{bma}}{4\pi} \sigma_{\text{tot}}^{mp}(\hat
s_1){\hat K}_{ab},\end{equation} where the quantities
\begin{equation}{\hat K}_{ab}\,=\,\int d{x_F}\int dk^2_{\bot}\,K_{ab}(x_F,k^2_{\bot})\end{equation}
 do not depend on energy, and $\sigma_{\text{tot}}^{mp}(\hat
s_1)$ is a constant up to logarithmic and power corrections.

 For estimation we take the total cross sections
$\sigma_{\text{tot}}^{\bar K^0p}$ and $\sigma_{\text{tot}}^{K^+p}$
to be a constant ($\sigma_{\text{tot}}^{\bar
K^0p}\sim\sigma_{\text{tot}}^{K^+p}\sim 20~\text{mb}$ at
$\sqrt{s_1}\gtrsim 10~\text{GeV}$.)  Then we obtain for the
production cross sections\footnote{The values in (\ref{numerics})
correspond to the region $x_F \gtrsim 0$. For $pp$ scattering the
total cross sections are two times larger.}

\begin{equation}\label{numerics}
\begin{gathered}
\sigma(pp\to \Theta^+(1540)X)=
0.8\,(1.6)\times\frac{\Gamma_{\Theta\to
K^0p}}{1~\text{MeV}}~\mu\mathrm{b},\\ \sigma(pp\to
\Lambda^+(1520)X)=106\,(126)~\mu\mathrm{b},
\end{gathered}
\end{equation} where the first values refer to the form factor (A) and
the second ones to the form factor (B).
The result for $\sigma(pp\to \Theta^+X)$ matches well that of Ref.
\cite{Vera} for the inclusive $pp\to\Theta^+X$
production\footnote{The authors of Ref. \cite{Vera} used the
standard expression for the inclusive cross section  and
$\sigma_{\text{tot}}$ \cite{Yao} taken from the parametrization of
experimental data at low energies.} at $\sqrt{s}\lesssim 10$~GeV.
If $\Gamma_{\Theta KN}=0.36\pm 0.11$~MeV as is claimed in
\cite{diana}, our result for the $\Theta^+$ production cross
section should be correspondingly smaller.

In scattering hadronic probes at high energy from nuclear target
the only positive signal for the $\Theta^+$ decaying to $K_S^0p$
 was reported by the SVD Collaboration, using 70 GeV proton in
 a fixed target arrangement $pA\to \Theta^+X$ at a center-of-mass energy of
 about 11.5 GeV.
 Their initial report \cite{SVD-1} recently was supported by a more detailed
 analysis \cite{SVD-2}
 which increased their pentaquark signal by a factor of about 8.
 Our prediction for $\sigma(pp\to \Lambda(1520)^+X)$ agrees with
the preliminary result of the SVD-2 collaboration \cite{SVD-2},
but $\sigma(pp\to \Theta^+X)$ is lower than the preliminary cross
section estimation (for $x_F>0$) of Ref. \cite{SVD-2}:
$\sigma\cdot \text{Br}(\Theta^+\to pK^0)\sim 6~\mu$b.

The illustrative examples of $x_F$ and $k_{\bot}$ distributions
for the $\Theta^+$ and $\Lambda(1520)$ are shown in Figs.
\ref{fig:pp_inclusive:x}, \ref{fig:pp_inclusive:k}  for the form
factor A. For the average transfer momenta squared of the
$\Theta^+$ we get
\begin{equation} \langle k_{\bot}^2\rangle=0.29\,(0.17)~\text{GeV}^2, \end{equation}
while for the $\Lambda(1520)$
\begin{equation} \langle k_{\bot}^2\rangle=0.73\,(0.22)~\text{GeV}^2, \end{equation}
respectively, where as above the first values refer to the form
factor (A) and the second ones to the form factor (B).

The ratio of $\Theta^+$ to $\Lambda(1520)$ production
cross-sections is $\sim 1\%$. Our estimation is a bit larger than
that obtained in the fragmentation-recombination model
 \cite{titov} but still is rather small and probably can be useful to explain why the
$\Theta^+$ production is suppressed in some high energy
experiments.

\begin{figure}
\includegraphics[width=50mm,keepaspectratio=true]{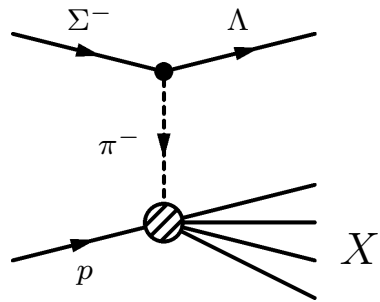}
\caption{The $\pi$-meson exchange diagram in $\Lambda(1520)$
production in inclusive $\Sigma p$ scattering
\label{fig:sigma_lambda}}
\end{figure}

In the same way it is possible to make quantitative predictions
for other type of colliding particles. As an example, we estimate
the cross section for the inclusive $\Lambda(1520)$ production in
$\Sigma^- p\to\Lambda(1520)$ collisions  at 600 GeV/c studied in
the fixed target Fermilab experiment E771 (SELEX). In the
fragmentation region of the $\Sigma$-hyperon this reaction can
proceed via the $\pi$-meson exchange, see
Fig.~\ref{fig:sigma_lambda}. Using
$\frac{G_{\Lambda\pi\Sigma}^2}{4\pi}$ from (\ref{glspi}) and
$\sigma(\pi N)=25~\mu$b we get
\begin{equation}
\sigma(\Sigma p\to \Lambda(1520)X)= 314\,(340)~\mu\mathrm{b},
\end{equation}
where, as before, the first value refer to the form factor (A) and
the second ones to the form factor (B). Taking $\hat x=0.83$ from
Table \ref{table:results}  for the ratio of inclusive $\Sigma p$
and $pp$ cross sections we get
\begin{equation}
\frac{\sigma(\Sigma p\to\Lambda(1520)X)}{\sigma(p
p\to\Lambda(1520)X)}\approx 2.9\,(2.7),
\end{equation}
that agrees with the preliminary experimental result
\begin{equation}
\frac{\sigma(\Sigma p\to\Lambda(1520)X)}{\sigma(p
p\to\Lambda(1520)X)}\approx 2.6,
\end{equation}
of the SELEX collaboration \cite{SELEX}.

\section{Conclusions}

Let us recall that our estimations may somehow depend on specific
assumptions regarding for instance the $K$-meson exchange
dominance at forward direction, and on the choice of the form
factor. As an outlook, it would be interesting to go beyond the
present calculation and to perform a systematic study of $K$,
$K^*$ and $\pi$ Regge exchanges into inclusive production of
(anti)strange baryons in $pp$ collisions. We plan to come back to
these issues in a next publication.

\begin{acknowledgments}
We are grateful to K.G.Boreskov, A.G.Dolgolenko, B.L.Ioffe, and
A.B.Kai\-dalov for the discussions. This work was supported by
RFBR grants 04-02-17263, 05-02-17869,  06-02-17120, and by the
grant for leading scientific schools 843.2006.2 .
\end{acknowledgments}

\begin{table}
\caption{The production cross sections (in units of $\mu$b) and
$\langle k_\perp^2\rangle$ (in units of GeV$^2$). Also are shown
the values of $\hat x$ explained in the text
\label{table:results}}
\begin{tabular}{|c|c|c|c|c|c|c|c|}
\hline & & \multicolumn{3}{|c|}{$\Theta^+(1540)$} &
\multicolumn{3}{|c|}{$\Lambda(1520)$} \\
 \hline& & $\sigma$ & $\langle k_\perp^2\rangle$ & $\hat x$ & $\sigma$
& $\langle k_\perp^2\rangle$ & $\hat x$  \\
\hline\hline $pp$ & A & 0.8 & 0.29 & 0.90 & 107
& 0.73 & 0.83\\ \hline & B & 1.56 & 0.17 & 0.91 & 126 & 0.22 & 0.88  \\
\hline\hline $\Sigma p$ & A & & & & 314 & 0.60 & 0.83 \\
\hline & B & & & & 340 & 0.21 & 0.85 \\ \hline
\end{tabular}
\end{table}


\begin{thebibliography}{9}

\bibitem{jlab} For
the latest reviews see ``Experimental Review of Pentaquarks'' by
A.~Sandorfi and ``Experimental summary'' by P.~Stoler in
Proceedings of the Pentaquark 2005 Workshop, Jlab, Newport News,
VA USA, October 20-22, 2005.
\bibitem{diana} V.V.~Barmin et al., hep-ex/0603017.
\bibitem{HERA-B}
I.~Abt et al.  [HERA-B Collaboration], Phys. Rev. Lett.
\textbf{93}, 212003 (2004).
\bibitem{SVD-1} A.~Aleev et al., hep-ex/0401024.
\bibitem{SVD-2} A.~Aleev et al., hep-ex/0509033.
\bibitem{Yao} T.~Yao, Phys. Rev. \textbf{125}, 1048 (1962).
\bibitem{halla}
JLab Hall A experiment E-04-012.
\bibitem{Cahn:2003wq}
R.N.~Cahn and G.H.~Trilling, Phys.~Rev.~ D \textbf{69}, 011501
(2004); hep-ph/0311245.
\bibitem{piling} V.M.~Belyaev and B.L.~Ioffe, Sov.~Phys.~JETP \textbf{56}, 493
(1982); T.~Pilling, Int.~J.~Mod.~Phys. A~\textbf{20}, 2715 (2005);
hep-th/0404131.
\bibitem{Vera} V.Yu.~Grishina et al., Eur.~Phys.~J.~A \textbf{25},
141 (2005).
\bibitem{titov} A.I.~Titov et al.,  Phys. Rev. C \textbf{70}, 042202
(2004).
\bibitem{SELEX} A.G.~Dolgolenko and V.A.~Matveev (private communication).

\end{thebibliography}
\end{document}